\documentclass[twocolumn]{aastex63}
\accepted{December 30, 2020}
\received{October 19, 2020}
\revised{December 20, 2020}
\submitjournal{ApJ}

\shorttitle{}

\usepackage{color}		
\definecolor{midgray}{gray}{0.4}	
\definecolor{orange}{rgb}{1,0.5,0} 
\definecolor{blue}{rgb}{0,0,0.6}  
\usepackage{apjfonts}

\definecolor{ao}{rgb}{0.0, 0.0, 1.0}

\usepackage{scrextend}
\deffootnote[0.5em]{0em}{1em}{\textsuperscript{\thefootnotemark}\,}
\usepackage{etoolbox}
\makeatletter
\patchcmd{\NAT@citex}
  {\@citea\NAT@hyper@{\NAT@nmfmt{\NAT@nm}\NAT@date}}
  {\@citea\NAT@nmfmt{\NAT@nm}\NAT@hyper@{\NAT@date}}
  {}% Do nothing if patch works
  {}% Do nothing if patch fails
\patchcmd{\NAT@citex}
  {\@citea\NAT@hyper@{%
     \NAT@nmfmt{\NAT@nm}%
     \hyper@natlinkbreak{\NAT@aysep\NAT@spacechar}{\@citeb\@extra@b@citeb}%
     \NAT@date}}
  {\@citea\NAT@nmfmt{\NAT@nm}%
   \NAT@aysep\NAT@spacechar%
   \NAT@hyper@{\NAT@date}}
  {}% Do nothing if patch works
  {}% Do nothing if patch fails
\patchcmd{\NAT@citex}
  {\@citea\NAT@hyper@{%
     \NAT@nmfmt{\NAT@nm}%
     \hyper@natlinkbreak{\NAT@spacechar\NAT@@open\if*#1*\else#1\NAT@spacechar\fi}%
       {\@citeb\@extra@b@citeb}%
     \NAT@date}}
  {\@citea\NAT@nmfmt{\NAT@nm}%
   \NAT@spacechar\NAT@@open\if*#1*\else#1\NAT@spacechar\fi%
   \NAT@hyper@{\NAT@date}}
  {}% Do nothing if patch works
  {}% Do nothing if patch fails
\makeatother

\makeatletter
\def\blfootnote{\xdef\@thefnmark{}\@footnotetext}
\makeatother

\newcommand{\myemail}{tmorishita@stsci.edu}
\newcommand{\simgt}{\,\rlap{\lower 3.5 pt \hbox{$\mathchar \sim$}} \raise
1pt \hbox {$>$}\,}
\newcommand{\simlt}{\,\rlap{\lower 3.5 pt \hbox{$\mathchar \sim$}} \raise
1pt \hbox {$<$}\,}
\newcommand{\Msun}{M_{\odot}}

\newcommand{\logm}{\log M_*/\Msun}
\newcommand{\logZ}{\log Z_*/Z_\odot}

\newcommand{\id}{GDS24569}

\newcommand{\kms}{{\rm km~s^{-1}}}

\newcommand{\ci}{[\textrm{C}~\textsc{i}]}

\newcommand{\gf}{5} % gas mass fraction
\newcommand{\gff}{5} % gas mass fraction
\newcommand{\gm}{7.1\times10^{9}\,M_\odot}

\newcommand{\snten}{4.1}

\newcommand{\hst}{{HST}}

\newcommand{\spit}{{Spitzer}}

\newcommand{\Z}{{\rm [Z/H]}}

\newcommand{\affilA}{Space Telescope Science Institute, 3700 San Martin Drive, Baltimore, MD 21218, USA; \href{mailto:\myemail}{\myemail}}
\newcommand{\affilB}{INAF/IRA, Istituto di Radioastronomia, Via Piero Gobetti 101, 40129, Bologna, Italy}
\newcommand{\affilC}{Dipartimento di Fisica e Astronomia dell'Universit\`a degli Studi di Bologna, via P. Gobetti 93/2, 40129 Bologna, Italy}
\newcommand{\affilD}{The Carnegie Observatories, 813 Santa Barbara Street, Pasadena, CA 91101, USA}
\newcommand{\affilE}{Institute of Astronomy and Astrophysics, Academia Sinica, Taipei 10617, Taiwan}

\begin{document}
\title{ \Large
Extremely Low Molecular Gas Content in the Vicinity of a Red Nugget Galaxy at $z=1.91$
%In search of gas content in the vicinity of a red nugget galaxy at $z=1.91$
}

\author{T.~Morishita}
\affiliation{\rm \affilA}

\author{Q.~D'Amato}
\affiliation{\rm \affilB}
\affiliation{\rm \affilC}

\author{L.~E.~Abramson}
\affiliation{\rm \affilD}

\author{Abdurro'uf}
\affiliation{\rm \affilE}

\author{M.~Stiavelli}
\affiliation{\rm \affilA}

\author{R.~A.~Lucas}
\affiliation{\rm \affilA}

\submitjournal{ApJ}

% ======================================================================
\begin{abstract}
We present Atacama Large Millimeter/submillimeter Array (ALMA)  Band 5 observations of a galaxy at $z=1.91$, \id, in search of molecular gas in its vicinity via the \ci~$^3$P$_1$-$^3$P$_0$ line. \id\ is a massive ($\logm=11$), passively evolving galaxy, and characterized by compact morphology with an effective radius of $\sim0.5$\,kpc. 
We apply two blind detection algorithms to the spectral data cubes, and find no promising detection in or around \id\ out to projected distance of $\sim320$\,kpc, while a narrow tentative line ($\snten\sigma$) is identified at $+1200$\,km\,/\,s by one of the algorithms. From the non-detection of \ci, we place a $3\sigma$ upper limit on molecular hydrogen mass, $\sim \gm$, which converts to an extremely low gas-to-stellar mass fraction of $\simlt\gf \%$. We conduct a spectral energy distribution modeling by including optical-to-far-infrared data, and find a considerably high ($\sim0.1\%$) dust-to-stellar mass ratio, i.e. $\sim10$-$100\times$ higher than those of local early-type galaxies. In combination with a previous result of an insufficient number of surrounding satellite galaxies, it is suggested that \id\ is unlikely to experience significant size evolution via satellite mergers.
We discuss possible physical mechanisms that quenched \id.
\end{abstract}


% =====================================================================

\section{Introduction}\label{sec:intro}
The origin of the most massive galaxies in the local universe has been a long-standing subject of galaxy evolution. From analysis of stellar populations and chemical abundances of local galaxies, it has been inferred that the most massive galaxies in the local universe completed their formation at redshift $z\simgt2$ \citep{kauffmann03,thomas03,gallazzi05,treu05,mcdermid15}. Recent near-infrared (NIR) observations indeed found many distant galaxies to be already quenched, showing consistency with the quenching timeline inferred from the local and low-$z$ universe \citep{brammer09,marchesini09,muzzin13,tomczak14}.

Interestingly, a large fraction of high-$z$ massive quenched galaxies are characterized by compact morphology \citep{cimatti04,daddi05,trujillo07,buitrago08,ichikawa10,cassata11}. While the average size of galaxies is smaller at that high redshift \citep{trujillo06,morishita14,vanderwel14}, high-resolution imaging by the Hubble Space Telescope (\hst) and the adaptive optics of ground-based facilities revealed that some of them have even smaller radius compared to the local counterpart at a similar mass, by a factor of $\sim5$ \citep{vandokkum08,damjanov09,szomoru10}. Given the absence of such a compact galaxy population in normal fields in the local universe \citep[][but see \citealt{valentinuzzi10} for their presence in dense fields]{shen03,taylor10}, the transition of galaxies in the size-mass plane implies that many high-$z$ compact galaxies would have to experience significant size evolution.

Various scenarios of significant size evolution for compact galaxies have been proposed in the past decade, including dry/wet, major/minor mergers \citep{khochfar06,naab07,hopkins09,naab09,nipoti09} and AGN feedback \citep{fan08,damjanov09} \citep[see][for a thorough review]{conselice14}. Among these scenarios, the dry minor merger scenario has been popularly discussed and taken as the most successful scenario in terms of its efficiency of size increase; surrounding satellite galaxies accrete to the central compact galaxy, without disturbing the central core, and in this way it can efficiently evolve effective radii with a small increase of stellar mass \citep{naab09,hopkins09,oser10,trujillo11}, while not all of them may have to follow the same path \citep{nipoti12,newman12}. The resulting stellar populations and mass profiles from this inside-out evolution are in good agreement with findings at intermediate redshifts \citep{vandokkum10,patel13,morishita15,papovich15} and for local massive galaxies \citep{belfiore17,ellison18}.

However, such an interpretation remains indirect, and the data still admit other scenarios. For example, such compact galaxies may not have to be the typical massive galaxy population at low redshift. In fact, there are a significant number of compact galaxies but more preferably in cluster environments \citep{valentinuzzi10,poggianti13}. It has also been proposed that the observed trend of average galaxy sizes reflects progenitor bias \citep[e.g.,][]{carollo13,fagioli16}, where large radius galaxies may appear at later time and drive apparent size evolution while high-$z$ compact galaxies remain as they are. 

Toward more direct understanding of their following evolutionary path, \citet{marmol12} studied satellite galaxies around massive galaxies to calculate possible size increase that would likely happen through accreting these satellite galaxies. Interestingly, their results indicated that the extant satellite galaxies are not enough to account for significant size evolution to the local relation. Furthermore, \citet{morishita16} studied a compact galaxy at $z=1.91$, \id, in deep \hst\ images from the extreme deep field (XDF) project \citep[][]{illingworth16}, to search for satellite galaxies around it down to $\logm\sim7.2$. Their conclusion is that the number of satellites is not enough either, and extra mass increase by, e.g. additional star formation, is required for \id\ to be on the local size-mass relation. 

This poses the question whether additional in situ star formation is possible in such compact quenched galaxies. One missing key component in these studies is gas, in particular in the form of molecular hydrogen. While the stellar component is not enough for the mass increase required to be on the local relation, additional star formation caused by remaining or newly accreted gas could change their sizes significantly \citep[i.e. the scenario C in][]{morishita16}. 

To answer the question, we here present new observations with the Atacama Large Millimeter/submillimeter Array (ALMA) on \id. Our configuration allows wide field-of-view (FoV) coverage to the extent of its virial radius, $r\sim300$\,kpc. Due to the spectral window available for the source redshift, we target atomic carbon, \ci~$^3$P$_1$-$^3$P$_0$ (rest-frame frequency 492\,GHz, hereafter \ci), which is known as a good alternative tracer of molecular hydrogen in extragalactic systems, originated from photodissociation regions {\citep[e.g.,][]{papadopoulos04,bell07,offner14,salak19}. The line has an excitation temperature of $23.6$\,K and a critical density for collisions with hydrogen atoms of $\sim1.2\times10^3$\,cm$^{-3}$. While CO(4-3) is also available in the same spectral window, it is not optimal to search for cool gas owing to its slightly higher excitation temperature (55.3\,K). Atomic carbon is found to well trace CO, at least in low-density environments i.e. at the boundaries of molecular clouds \citep{glover15}, at an almost constant ratio of $N(\ci)/N({\rm CO})\sim0.1-0.2$ in the local universe \citep{keene97,ojha01,ikeda02}. Early studies presented \ci\ as an excellent gas tracer at high redshifts too \citep[e.g.,][]{weiss05,walter11,valentino18}. 
%reported the tight relation at $z\sim1.2$.
}

In this study, we exploit the exquisite sensitivity of ALMA and the ubiquity of \ci\ to search for previously unseen gas content in and around \id, aiming to infer the primary quenching mechanism and possible future evolutionary path. The paper is structured as follows: In Section~\ref{sec:data}, we describe our observations and data reduction. In Section~\ref{sec:method}, we present our analysis method of a blind search for gas in the ALMA data, and results. While the primary goal in this study is to search for surrounding gas, the dataset also provides us an opportunity to investigate the cause of quenching in the central galaxy from a panchromatic analysis over optical-to-far-infrared (FIR) wavelengths (Section~\ref{sec:panc}).  In Section~\ref{sec:dis}, we discuss our results and present our interpretation. Throughout, magnitudes are quoted in the AB system assuming $\Omega_m=0.3$, $\Omega_\Lambda=0.7$, $H_0=70\,\kms\, {\rm Mpc}^{-1}$, and we assume the \citet{salpeter55} initial mass function.

%%%%%%%%%%%%%%%%%
\begin{figure*}
\centering
	\includegraphics[width=0.48\textwidth]{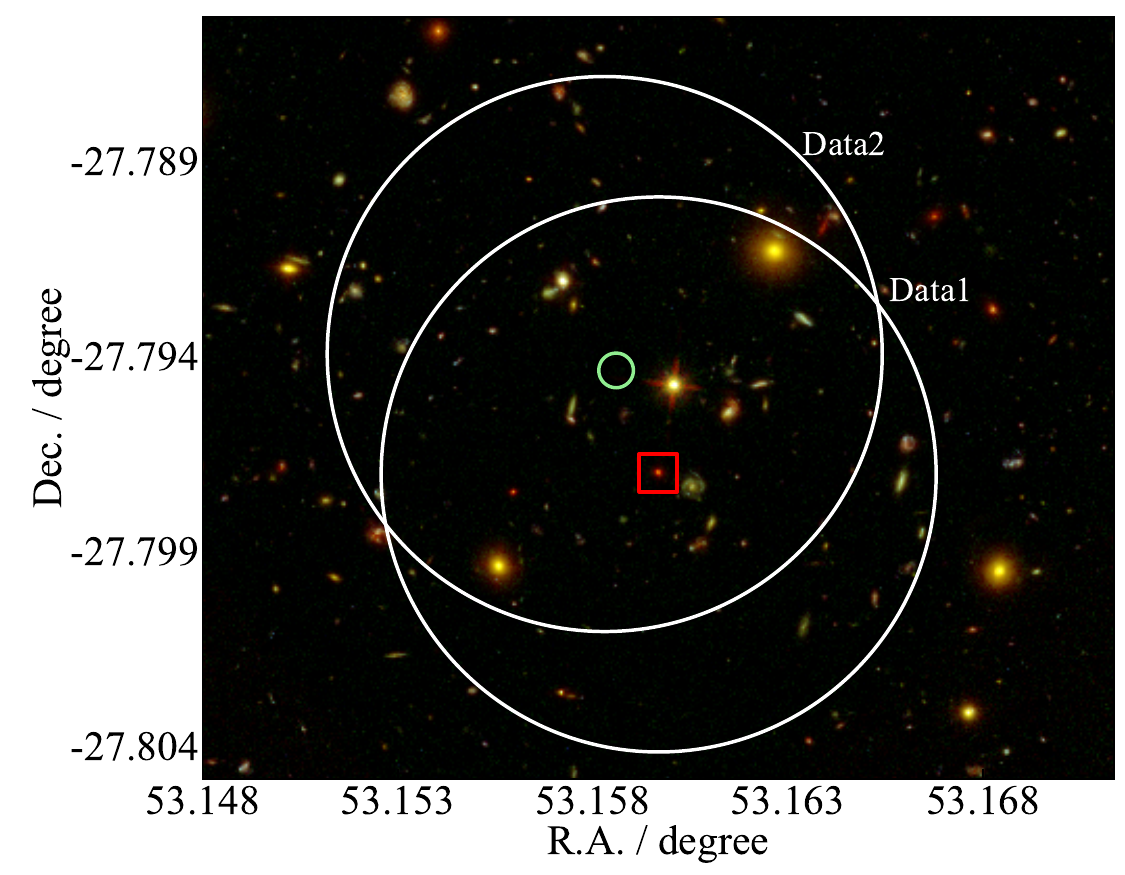}
	\includegraphics[width=0.42\textwidth]{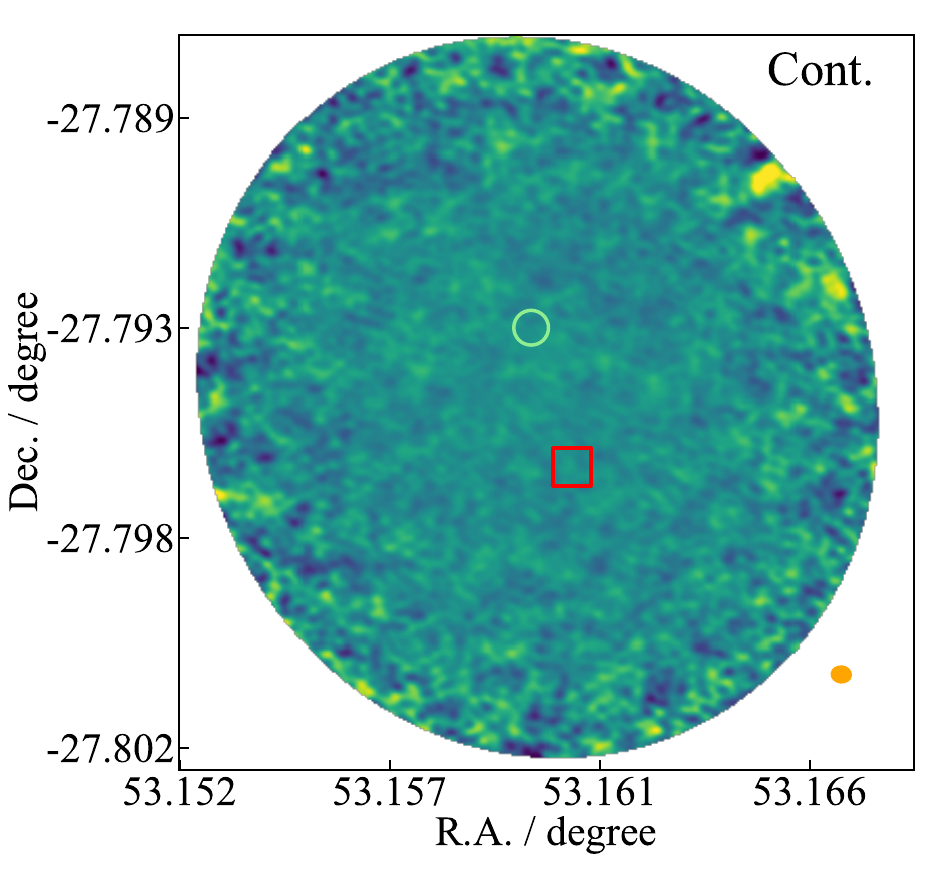}
	\caption{
	(Left): Pseudo RGB image (F160W+F814W+F435W of \hst) of the observed field. The two ALMA pointings are overlaid (Data1 and Data2; white circles). The target, \id, is centered in the image (red square). The position of the tentative detection found in Data2 (Section~\ref{sssec:damato}) is also shown (green circle), while no optical/IR counterpart is found at the position.
	(Right): Dust continuum map, created by collapsing the entire frequency range of the combined data cube. The coordinates of the objects in the left panel are indicated with the same symbols. No continuum emission is detected near the position of \id. The beam size is shown at the bottom right (orange ellipse).
	}
\label{fig:fov}
\end{figure*}

%%%%%%%%%%%%%%%%%
\section{Data and Analysis}\label{sec:data}

\subsection{Target galaxy}\label{ssec:reduction}
Our primary target, \id, is a massive ($\logm\sim11$) galaxy at $z=1.91$, originally reported in \citet{daddi05}, identified in a spectroscopic campaign with \hst, GRAPES \citep{pirzkal04}. The galaxy is passively evolving, characterized by its spectral features, and has compact morphology with effective light radius $\sim0.5$\,kpc \citep{szomoru10}, which converts to stellar mass density of $\log\Sigma_* / M_\odot {\rm kpc}^{-2} \sim11$ (Figure~\ref{fig:fov}). \citet{morishita19} analyzed its spectral energy distribution (SED) by fitting photometric and spectroscopic data points, and revealed that the galaxy experienced a very short period of star formation $\sim0.5$\,Gyr prior to its observed redshift, characterizing it as a quenched galaxy (see also Section~\ref{sec:dis}). 

On the size-mass plane, \id\ is located $\sim2\sigma$ below the median size of passive galaxies at that redshift and stellar mass \citep{vanderwel14}. To speculate its future evolutionary path, \citet{morishita16} investigated surrounding sources photometrically selected in extremely deep images from the XDF project \citep{illingworth13}, down to $\sim10^8\,\Msun$. There are insufficient satellite galaxies to move \id\ onto the local size mass relation when following a simple formula for a minor merger scenario, $\Delta r \propto \Delta M^2$ \citep{naab09}. While \id\ may end up as one of the compact population found in the local universe, it is worth investigating any contribution from optically-dark components, which may trigger star formation and size evolution, before we conclude its fate.

%%%%%%%%%%%%%%%%%
\subsection{ALMA Cycle 7 observation}\label{ssec:reduction}
Interferometric observations with Band 5 were executed in ALMA cycle 7 (PID. 2019.1.01127.S, PI T. Morishita). We designed the observations so they would reveal gas components down to $\sim10^8\Msun$ within the virial radius of \id, $\sim300$\,kpc, via the \ci\ emission line with an excitation temperature of $\sim30$\,K. Since our purpose here is not to resolve each gas clump, we optimize the antennas' configuration for a large FoV, resulting in the final beam size of $1.0\times\,0.8$\,arcsec$^2$. We set three 1.875\,GHz-width spectral windows, one of which was tuned to cover the \ci\ frequency of the target ($\sim169$\,GHz), while the other two spectral windows were centered on 168.5\,GHz and 170.4\,GHz for a continuum estimate.

Our observations, originally executed in November 2019 with on-source exposure of 107\,minutes, were accidentally off the central pointing for $\sim 12$\,arcsec, while \id\ was still within the FoV (dubbed as Data2). The observations were compensated in January 2020, with the target in the center as it was originally planned (Data1). We therefore acquired two datasets with the same setup for frequency and exposure time as described above (Figure~\ref{fig:fov}), which allow us an independent check on possible detection (Section~\ref{sec:method}). We ran the tclean task of CASA on each of the datasets, with parameters of pixel scale 0.16\,arcsec and velocity resolution element $\sim28$\,km\,/\,s ($\sim15.6$\,MHz).

%=================================
\section{Method and Results}\label{sec:method}

%%%%%%%%%%%%%%%%%
\subsection{Dust Continuum Emission}\label{ssec:dust}
In Figure~\ref{fig:fov}, we show a continuum map created by collapsing the spectral data cube over the entire velocity range, $-4000$-$3000$\,km\,/\,s. No continuum source is detected near the position of \id. Root-mean-squares (RMSs) are calculated by {\tt imstat} of CASA; 27\,$\mu$Jy\,/\,beam and 34\,$\mu$Jy\,/\,beam for Data1 and Data2, respectively, and 21\,$\mu$Jy\,/\,beam for the combined data cube at the target source position. This non-detection of dust emission is consistent with the optical-to-NIR SED analysis of \id, $A_V\sim0.4$\,mag \citep{morishita19}. We will use the upper limit derived here for a panchromatic analysis in Section~\ref{sec:dis}.

%%%%%%%%%%%%%%%%%%%%%
\begin{figure}
\centering
	\includegraphics[width=0.45\textwidth]{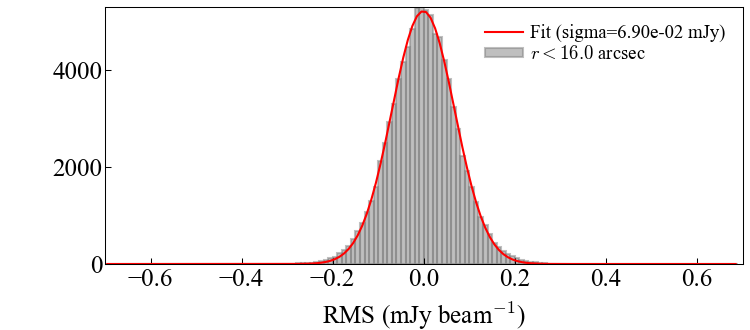}
	\includegraphics[width=0.48\textwidth]{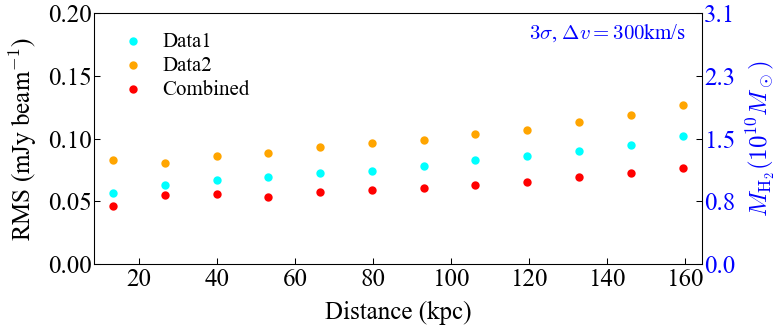}
	\caption{
	(Top): Distribution of fluxes measured at 1000\,random positions in one of the data cubes. Fluxes are in units of mJy per beam per spectral element.
	(Bottom): Radial distribution of RMSs estimated at each radius in the three data sets --- Data1 (cyan), Data2 (yellow), and combined (red) cubes. Corresponding molecular hydrogen mass limits ($3\sigma$) for a line width of $\Delta v=200$\,km\,/s are shown in the right $y$-axis.
	}
\label{fig:rand}
\end{figure}

%%%%%%%%%%%%%%%%%
\subsection{Clump Finding Algorithms}\label{ssec:cs}
Our primary goal in this study is to search for gas clumps in the data cubes via the \ci\ emission line, both in the position of \id\ and its neighboring regions out to $r\sim300$\,kpc, which is its approximate virial radius. Despite a complicated aspect of blind detection in interferometric data cubes, our advantage is that we have two independent data cubes, so that any detection found in one data cube can be checked in the other. We start with a fiducial detection algorithm, {\tt clumpdind} \citep{williams94}, but also apply a method introduced in \citet{damato20} to check the consistency.

%%%%%%%%%%%%%%%%%
\begin{figure}
\centering
	\includegraphics[width=0.48\textwidth]{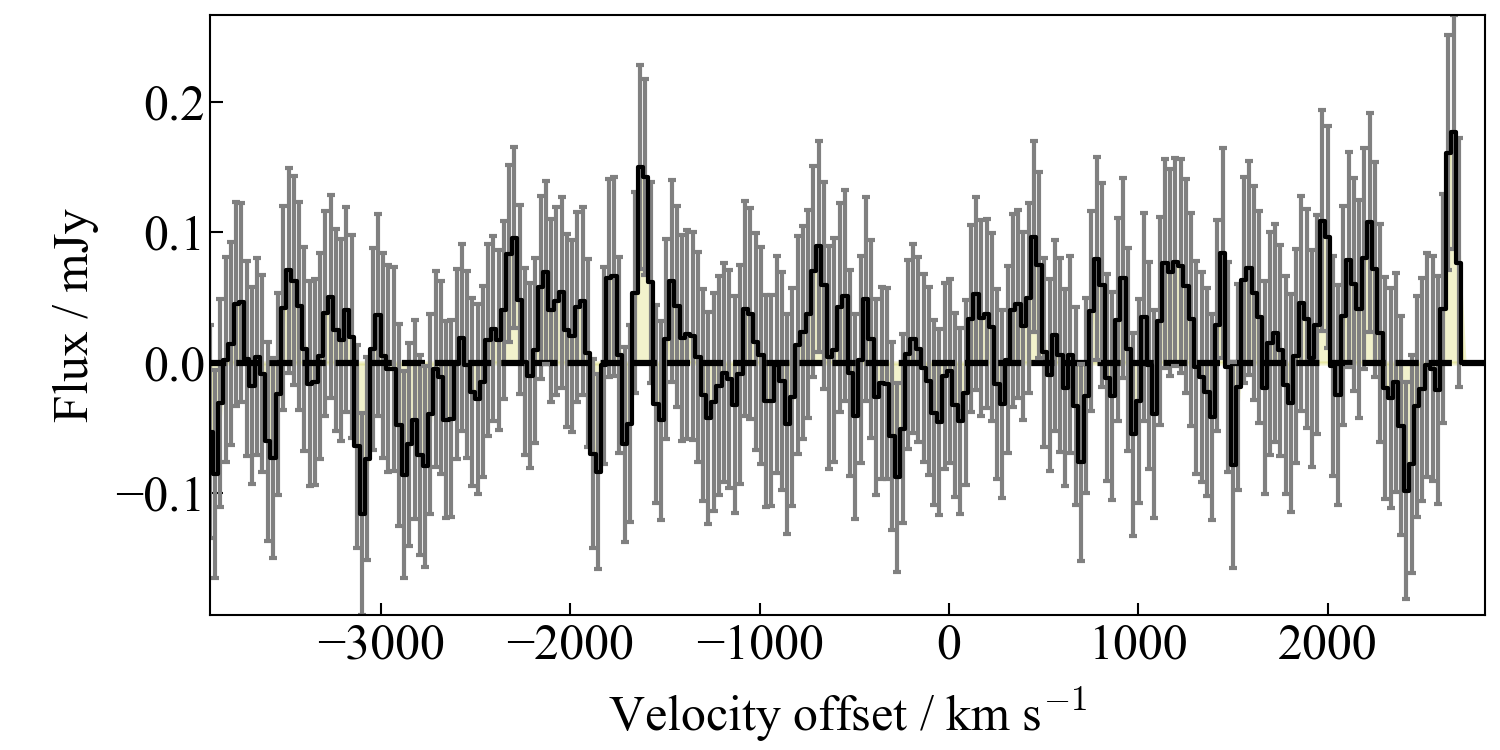}
	\caption{
	Spectrum extracted in the position of \id\ in the combined data cube, with an aperture of the beam size. No significant flux excess is seen.
	}
\label{fig:line_cen}
\end{figure}

%%%%%%%%%%%%%%%%%
\subsubsection{Starlink Clumpfind Algorithm}\label{sssec:cf}
To search for line detection across the data cubes, we apply an automated detection algorithm. We first run the {clumpfind} algorithm, implemented in the {\tt starlink} package \citep{currie14}. One of the parameters for clumpfind is the noise level of the data represented by RMSs. While {\tt imstat} of CASA provides a RMS value at each frequency value, the noise is correlated significantly for our case with a relatively large beam size. Furthermore, the data show differential noise structure along the radial direction from the center, where larger noise structures frequently appear at a larger radius. To estimate a more realistic noise level, we extract apertures of beam size in 1000 randomly selected positions in each cube, to measure RMSs in annuli at every 10\,pixels from the center. As an example, the distribution of extracted fluxes is shown in the top panel of Figure~\ref{fig:rand}, where the distribution is well fitted with a Gaussian. A radial distribution of derived RMSs ($=$\,FWHM\,/\,2.355 of Gaussian fit) are shown in the bottom panel of Figure~\ref{fig:rand}. The RMS varies from $\sigma\sim45$\,$\mu$Jy\,/\,beam\,/\,spectral element at the innermost region to 130\,$\mu$Jy\,/\,beam\,/\,spectral element at the outermost region.

We run the clumpfind algorithm on each of two data cubes with various RMSs taken from the estimated range above. Other configuration parameters are also determined through our initial tests (VELORES=5, MINPIX=50, and FWHMBEAM=3). With the setup, the algorithm initially detects $\sim100$ sources from the entire FoV of each dataset for different RMS values. However, most of the identified clumps are located at outer regions, $r\simgt15$\,arcsec. Our visual inspection does not confirm any of these clumps as real sources. Furthermore, none of these sources are commonly identified in both of the data cubes. We thus conclude that all of the detected clumps are artifacts. To confirm this, we also apply the same method to the data cubes but with signs of fluxes inverted. We find a similar number of detections with a similar spatial distribution as for the real cubes. 

The spectrum extracted from the position of \id\ in the combined cubes is shown in Figure~\ref{fig:line_cen}. No obvious line is detected beyond the noise limit associated with each spectral element.

%%%%%%%%%%%%%%%%%%%%%
\subsubsection{\citet{damato20}'s method}\label{sssec:damato}
To double-check the non-detection in Section~\ref{sssec:cf}, we also apply a blind detection method introduced in \citet{damato20},
that relies on
spectral, spatial and reliability criteria. 

We perform the detection on
the signal-to-noise-ratio- (S/N-) cube. In order to produce it, we
firstly generate a noise-cube, where the pixels have the value of the
RMS in a box centered in them. We choose a box with an area equal to
that of 10\,beams, since we find that this size allows us to trace the
local variation of the noise in each channel, still having enough
statistics per box. The RMS is recursively calculated to convergence
masking all pixels above $3\times$\,RMS at each iteration. The S/N-cube
is then obtained as the ratio between the original data-cube and the
noise cube. 

In order to perform the detection, we firstly scan each
spaxel searching for a given amount of contiguous channels
($\mathrm{N_{ch}}$) above a given S/N threshold. We search for
$\mathrm{N_{ch}}$=2, 3, 4, 5 (corresponding to $\sim$\,60, 90, 120, 150\,km\,/\,s, 
respectively) above a S/N threshold of 1.5, 2, 2.5 and 3. Then,
among the candidate detections, we reject all those with a number of
contiguous pixels lower than a given amount $\mathrm{N_{ch}}$. For each
combination of $\mathrm{N_{ch}}$ and S/N threshold, we try
$\mathrm{N_{px}}=2, 3, 4$. 

Finally, in order to estimate the incidence
of spurious detections, we perform the same algorithm on the
``negative" S/N-cube, where the pixels have the same value and inverted
sign of the original (``positive'') S/N-cube. Then, for any combination
of $\mathrm{N_{ch}}$, S/N threshold and  $\mathrm{N_{px}}$ we reject all
the candidate detections obtained from the positive S/N-cube that have a
peak S/N lower than the maximum peak S/N of the detections obtained from
the negative S/N-cube using the same parameter combination. Finally, we
crop the detection area to the half-primary-beam-width.

\begin{figure*}
\centering
	\includegraphics[width=0.35\textwidth]{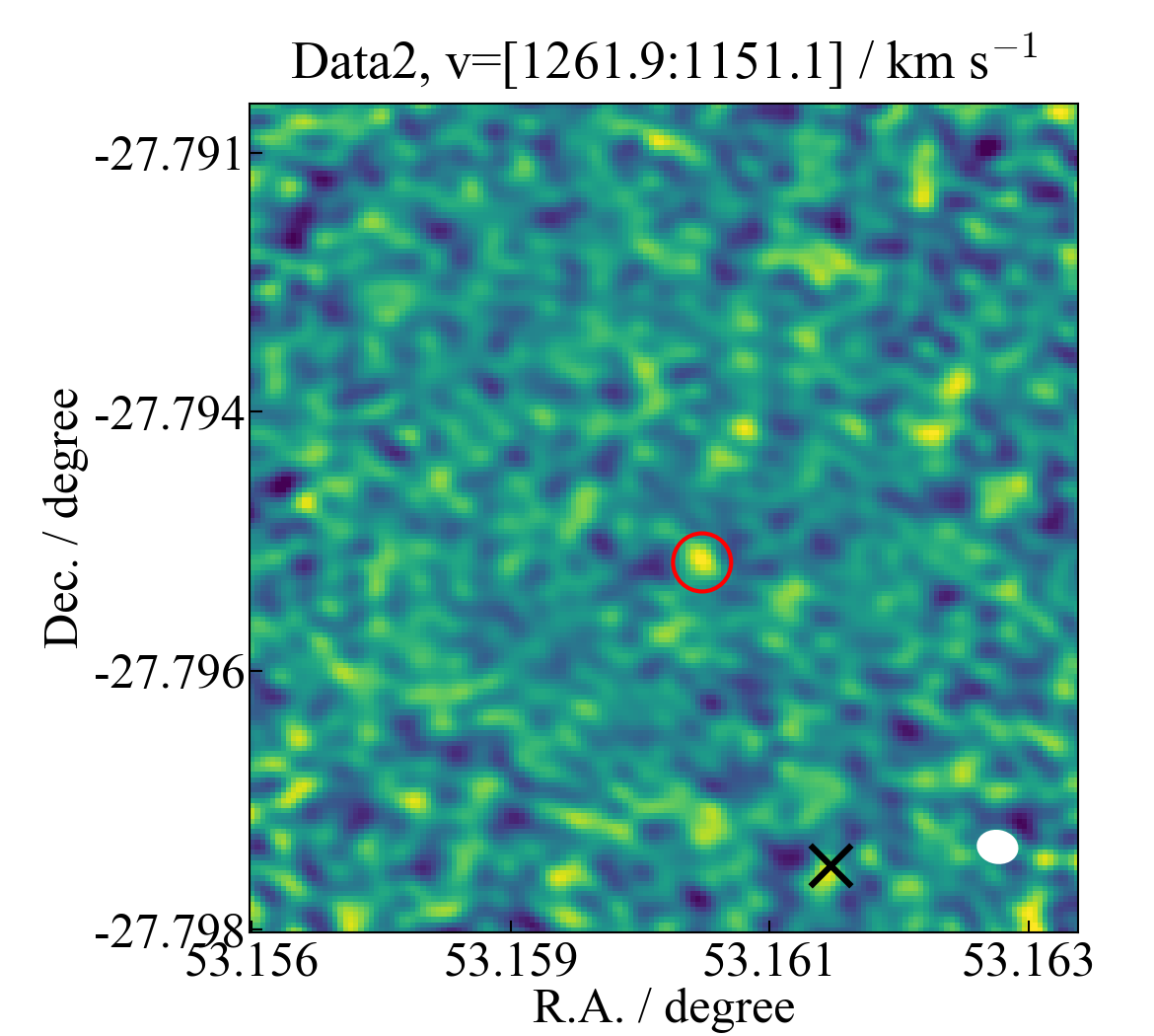}
	\includegraphics[width=0.6\textwidth]{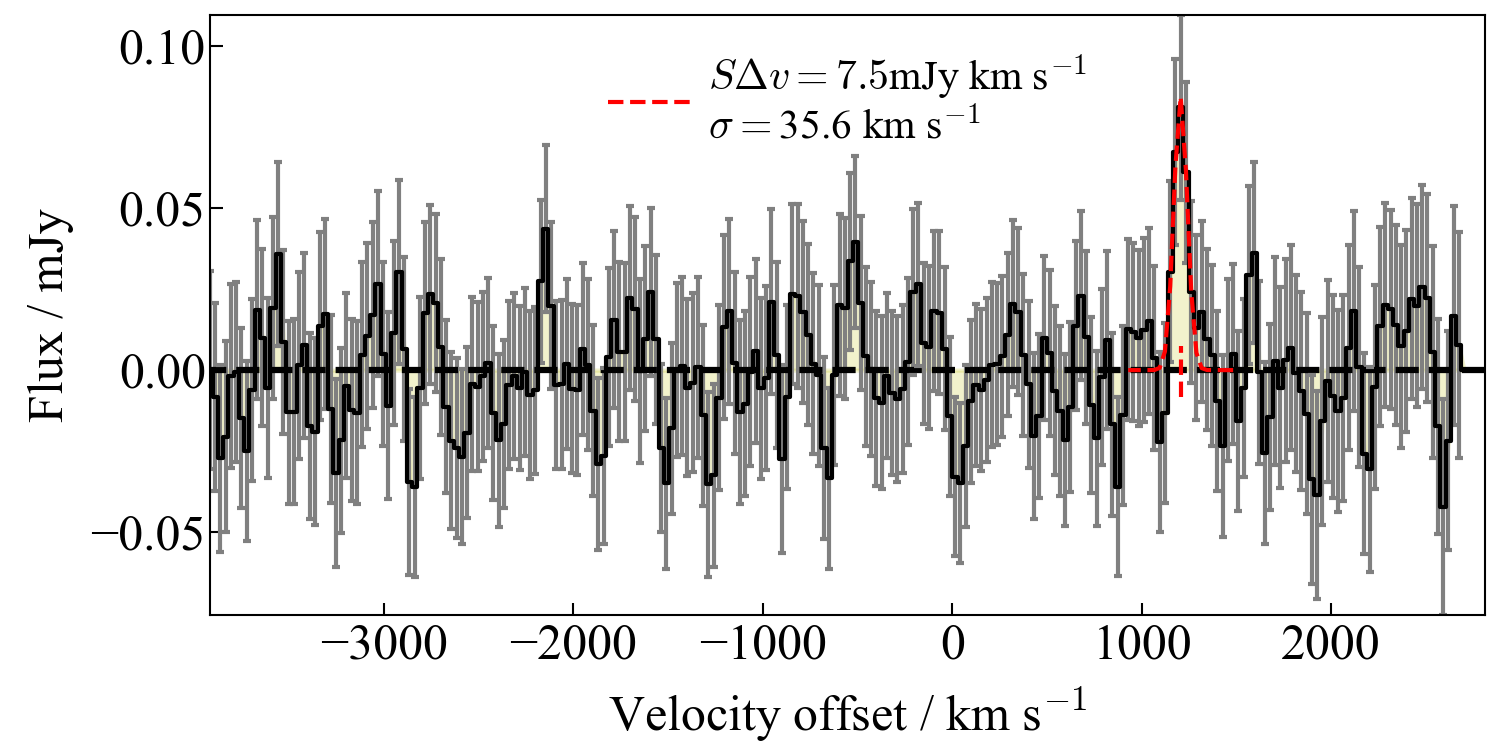}
	\includegraphics[width=0.35\textwidth]{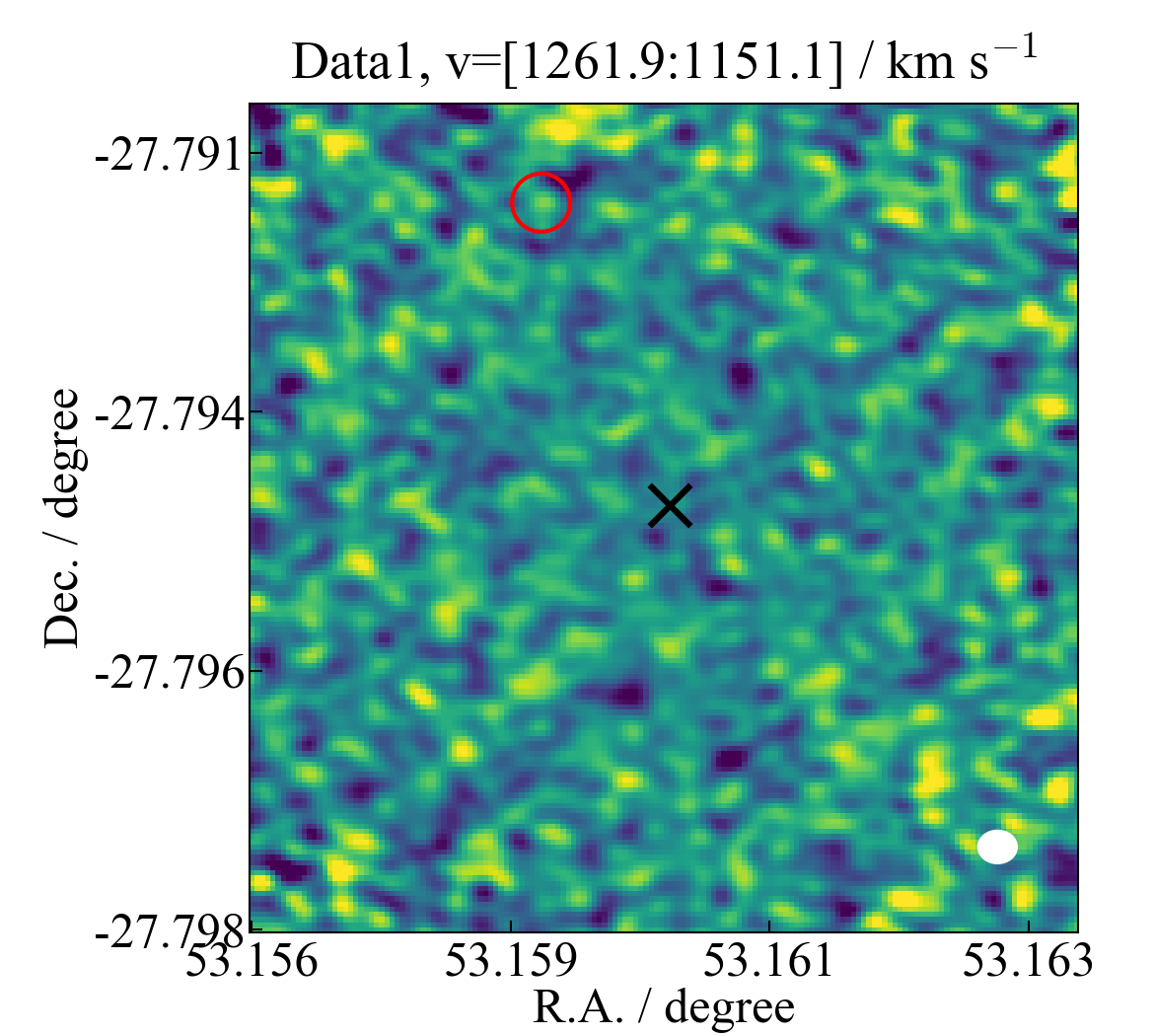}
	\includegraphics[width=0.6\textwidth]{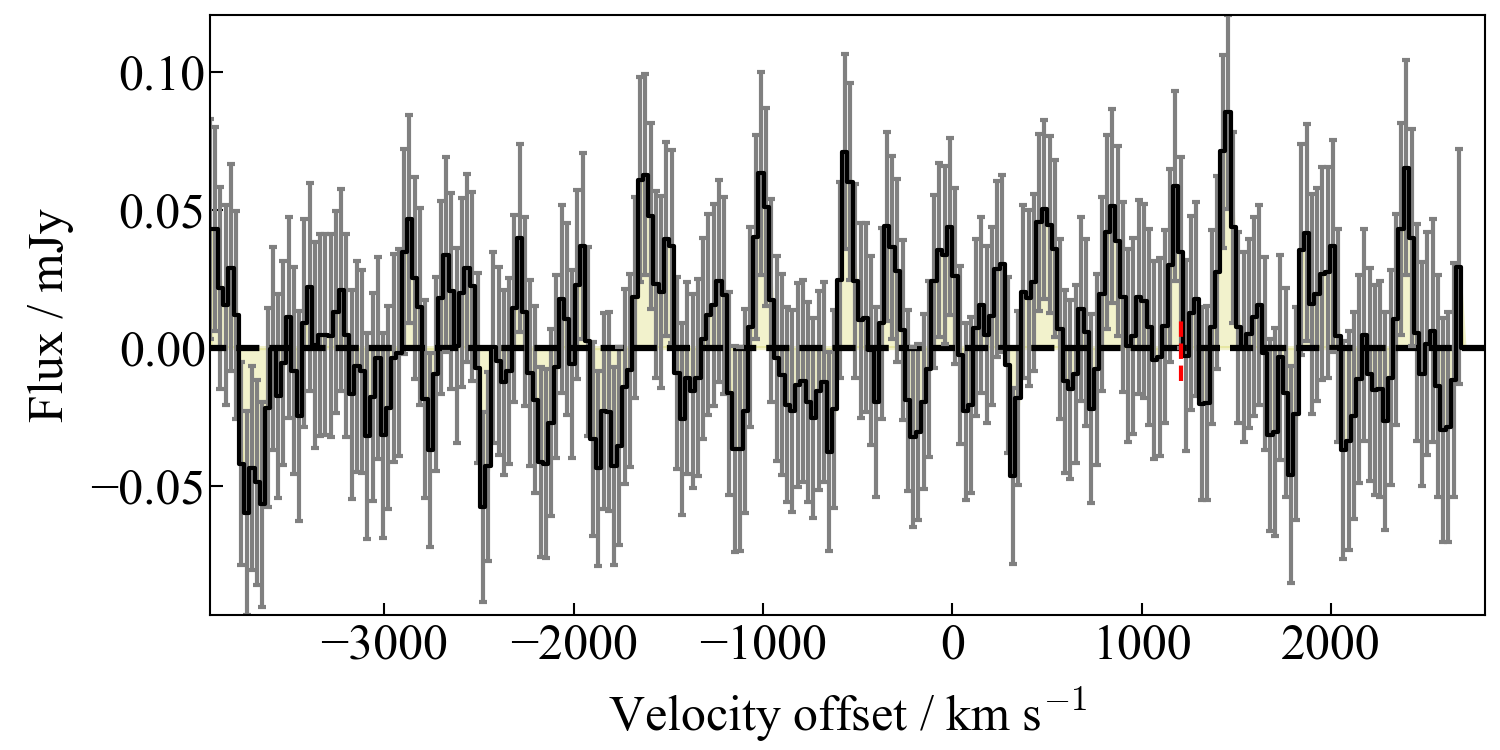}
	\caption{
	(Top): Tentative line detection at $V\sim1200$\,km\,/\,s in Data2 is shown (left panel; red circle) in the velocity integrated map (centered on its pointing). The position of \id\ is also shown (black cross). The beam size is shown at bottom-right (white ellipse). The Gaussian fit to the tentative line is overlaid in the extracted spectrum (right panel; red dashed line).
	(Bottom): Same as above but for Data1. No significant flux excess is seen at the velocity (red dashed line).
	}
\label{fig:line}
\end{figure*}

From this analysis, we find one tentative detection ($\snten\,\sigma$) at $+1200$\,km\,/\,s at projected distance $\sim14$\,kpc away from \id\ in Data2 (top panel of Figure~\ref{fig:line}), while no optical counterpart is identified in the XDF dataset (Figure~\ref{fig:fov}). {The RMS at the position is $87\,\mu$Jy\,/\,beam.} The line is narrow, and fitted with a gaussian of FWHM\,$=84$\,km\,/\,s ($\sim3$\,spectral elements). The total flux from the integral of the gaussian fit is $S_{\ci} \Delta v \sim 7.5$\,mJy\,km\,/\,s. 
{The line is not identified at the same spectral position in Data1, with only 1\,positive pixel of S\,/\,N\,$\sim1$ (bottom panel of Figure~\ref{fig:line}). This may be attributed to lower sensitivity at the position in Data1, which is $\sim10.5$\,arcsec away from the center, while the RMS value is $\sim120\,\mu$Jy\,/\,beam and the line should still be detectable at S\,/\,N\,$\sim3$. We thus leave this detection as tentative, while including this emission would not change our analysis and main conclusion in Section~\ref{sec:dis}.}

%%%%%%%%%%%%%%%%%%%%%

%%%%%%%%%%%%%%%%%%
\subsection{Upper limits in molecular gas mass}\label{sec:sample}
In this subsection, we attempt to place an upper limit on molecular hydrogen mass from the estimated RMSs in Section~\ref{sssec:cf}. Following a recipe by \citet[][also \citealt{wagg06,man19}]{papadopoulos04}, molecular hydrogen gas mass is inferred by:
\begin{eqnarray}
    M_{\rm H_2} = 1375.8 {D_l^2\over{(1+z)}} ({X\ci \over{10^{-5}}})^{-1}  \nonumber \\
    \times ({A_{10}\over{10^{-7}{\rm s^{-1}}}})^{-1} Q_{10}^{-1} {S_{{\ci}}\Delta v\over{\rm Jy\ km\ s^{-1}}} M_{\odot}.
\end{eqnarray}
The equation consists of two steps: conversion from observed \ci\ intensity to atomic carbon mass, and then to molecular hydrogen mass. The conversion involves a couple of parameters. We adopt the excitation factor $Q_{10}=0.5$ as a fiducial value, assuming local thermodynamic equilibrium (i.e. optically thin). $Q_{10}$ ranges from 0 to 1, depending on the temperature, density and radiation field, though none of these values can be constrained without line ratios to other excitation levels. {We adopt a molecular \ci-to-${\rm H_2}$ conversion factor $X{\ci}=1.5\times10^{-5}$ taken from \citep{jiao19}. The value is the lower limit measured in nearby galaxies, and gives us a conservative upper limit in molecular hydrogen mass, whereas the conversion factor can be as high as $X{\ci}=5\times10^{-5}$ \citep[e.g.,][]{frerking89,weiss03,weiss05}.} The conversion to molecular hydrogen mass includes the mass in Helium by a correction factor of 1.36 \citep[][]{solomon05}. $A_{10}=7.93\times10^{-8}$\,s$^{-1}$ is the Einstein A-coefficient, and $D_l$ is the luminosity distance to the source redshift, $z=1.91$.

Since our input for the equation above is the upper limit for flux density, we need to assume the line width to obtain a velocity-integrated line intensity, $S_{\ci} \Delta v$. 
{We set $\Delta v=300$\,km\,/\,s based on stellar velocity dispersion measurements in massive compact quiescent galaxies \citep[e.g.,][]{vandesande13,belli17}, while narrower line width has been seen in massive {\it rotating disks} \citep[$\sim200$\,km\,/\,s;][]{weiss05,man19}. With this we obtain $S_{\ci}\Delta v\sim42$\,mJy\,km\,/\,s\,/\,beam as a $3\sigma$ upper limit in the position of \id.}

By substituting these numbers in the equation above, we obtain a conservative $3\sigma$ upper limit on molecular hydrogen mass of $\sim\gm$, characterizing a significantly low gas mass fraction $f_{\rm gas}\simlt\gff\%$. 
%For the tentative line we found in Section~\ref{sssec:damato}, the same conversion provides $\sim1.3\times10^{9}\,M_\odot$. 
%{Note that while this value is slightly smaller than the detection limit in Figure~\ref{fig:rand}), this is due to the different algorithm used for rms measurement in the \citet{damato20}'s method.}
%{The value is smaller than the $3\sigma$ detection limit at this position (see the bottom panel in Figure~\ref{fig:rand}) due to its narrow line width.}

%%%%%%%%%%%%%%%%%%%%%
\begin{figure*}
\centering
	\includegraphics[width=0.6\textwidth]{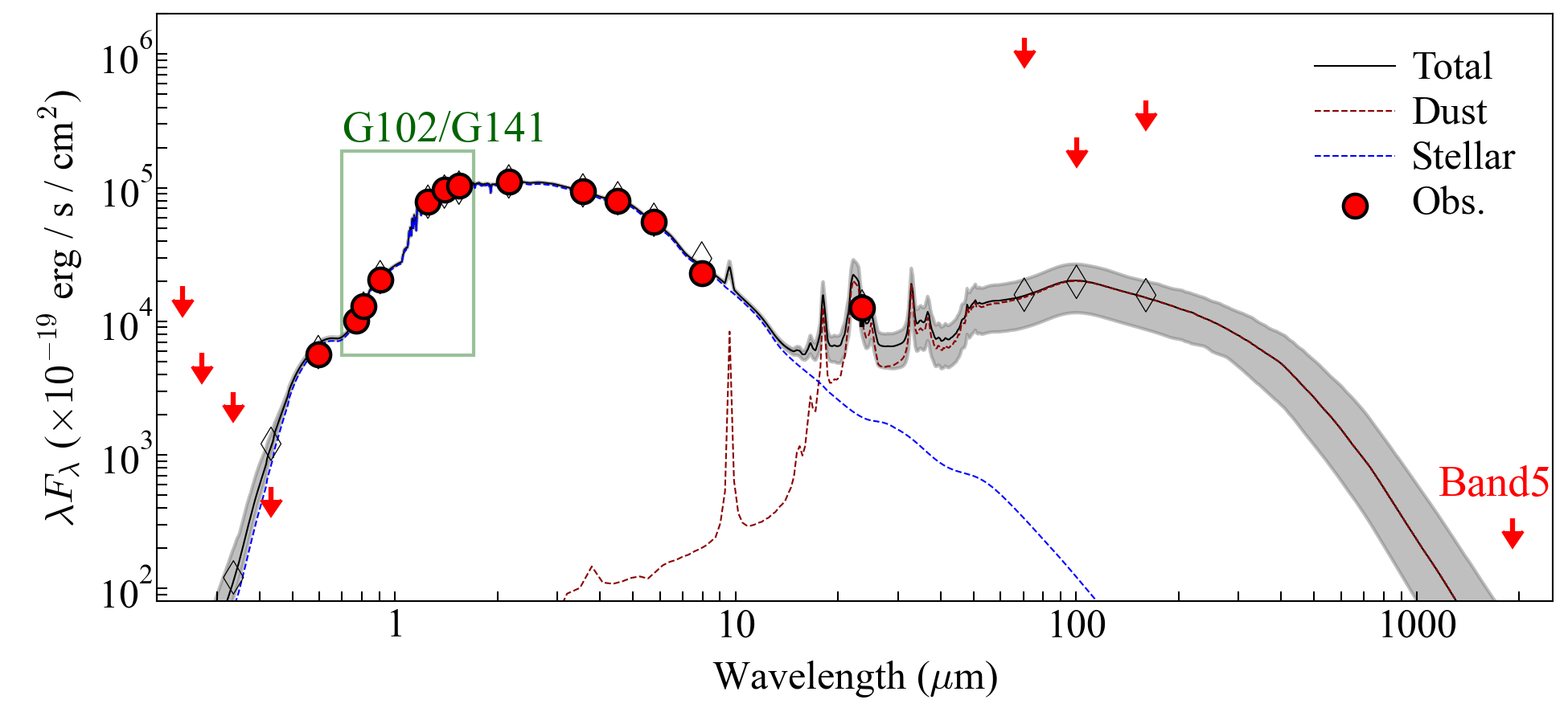}
	\includegraphics[width=0.3\textwidth]{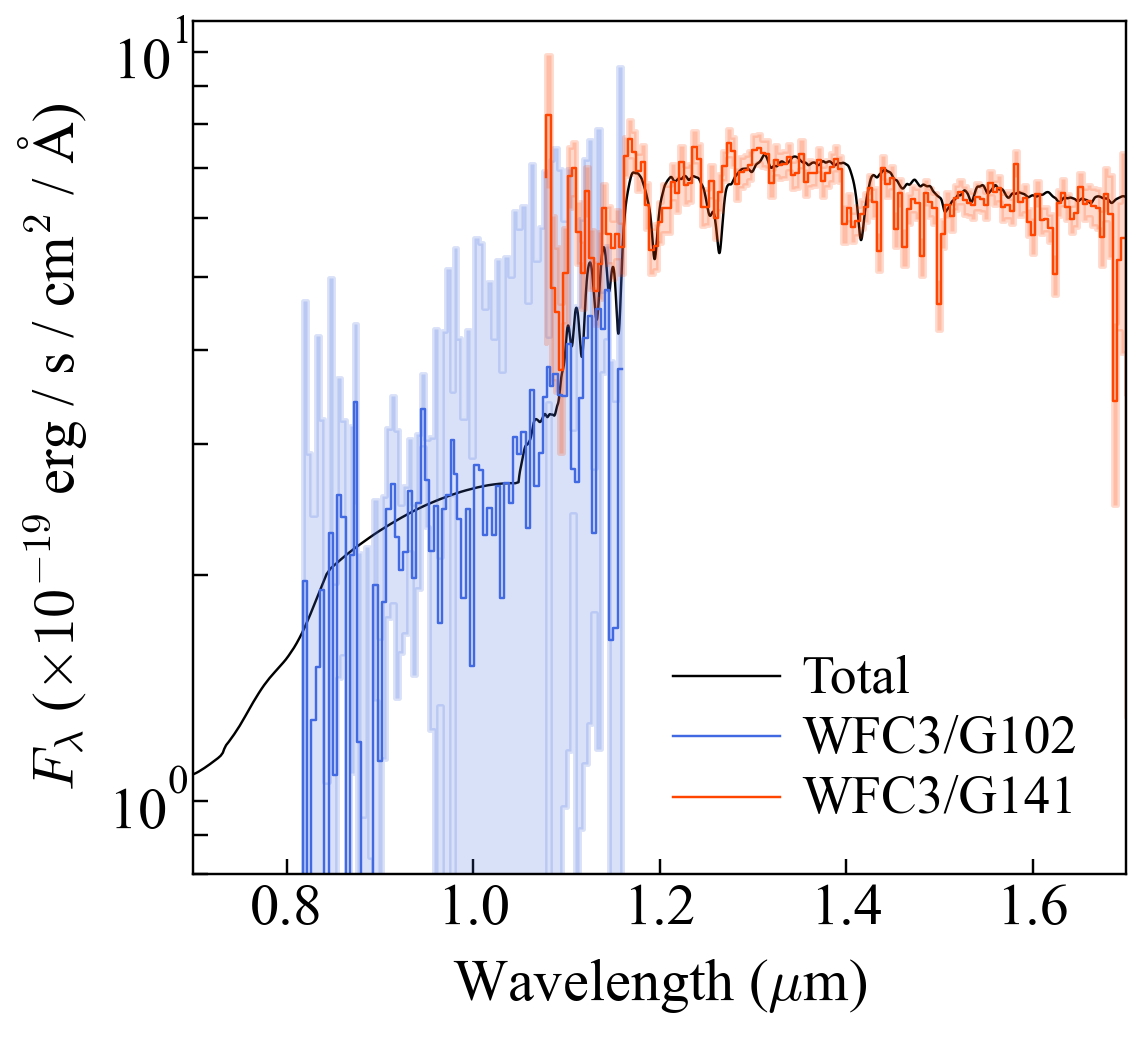}
	\includegraphics[width=0.9\textwidth]{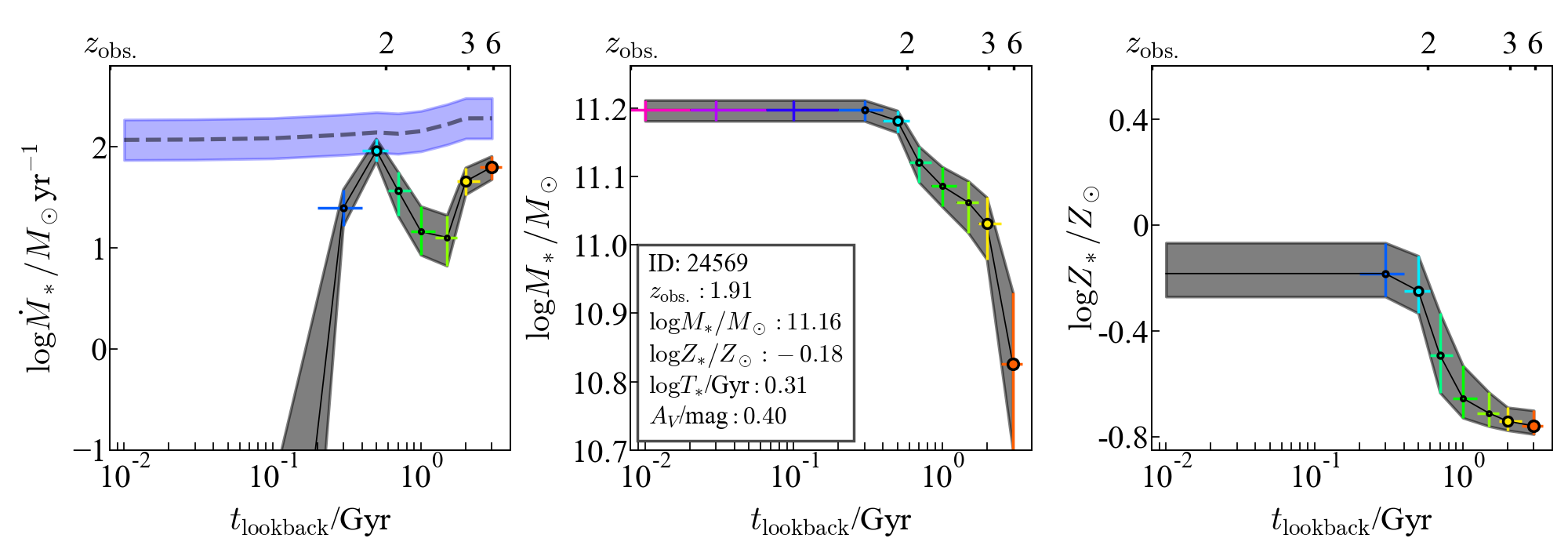}
	\caption{
	{
	(Top): Spectral energy distribution of \id. Broadband photometric data points used for the fit are shown (red points for detection and arrows for upper limits). The zoom-in region (green rectangle) is shown in the right panel to highlight the NIR grism spectra of WFC3/G141 and G102 (red and blue lines, respectively). Flux contribution from stellar and dust (blue and red dashed lines) components, as well as the best-fit total flux (black solid line with uncertainty shaded in gray).
	}
	(Bottom): Star formation (left), mass accumulation (middle), and metallicity evolution (right) histories of \id. 
	The star formation rate of the star-forming main sequence for a given stellar mass at each lookback time \citep{speagle14} is shown in the left panel (dashed line with blue hatch).
	}
\label{fig:sed}
\end{figure*}

%%%%%%%%%%%%%%%%%%
\subsection{Panchromatic analysis of \id}\label{sec:panc}
\citet{morishita19} investigated the star formation history of \id\ from a combined dataset of NIR spectrum and optical-to-NIR broadband photometry. We re-fit the dataset with a SED fitting code, {\tt gsf} (ver.1.4\footnote{\url{https://github.com/mtakahiro/gsf/tree/version1.4}}). From \citet{morishita19}, we retrieve broadband photometry of \hst\ and \spit\ up to IRAC CH4, originally published by the 3D-HST team \citep{skelton14}. Deep spectra of WFC3-IR G102 and G141 originally taken in 3D-HST, FIGS \citep{pirzkal17}, and CLEAR \citep{carpenter18}, which were reduced and presented in \citet[][]{morishita19}, are also included, to constrain its stellar populations and star formation history. 

{Specifically, in this redshift range, the deep grism spectra capture the Balmer break and absorption features at rest-frame $\sim4000$\,\AA, while the broadband coverage extending to rest-frame $K$-band is sensitive to variation in metallicity and dust attenuation. This comprehensive coverage enables to break the well-known age-dust-metallicity degeneracy, and enables robust characterization of star formation histories too. Interested readers are referred to \citet{morishita18,morishita19}, that presented intensive tests with simulated datasets, for more details.}

We include the upper limit on the continuum obtained from our ALMA observations (Section~\ref{ssec:dust}). We also include \spit\ MIPS $24\mu$m flux from \citet{whitaker14}, and upper limits on MIPS $70\mu$m, Herschel PACS $100$ and $160\,\mu$m from the GOODS Herschel program \citep{elbaz11}. The broadband photometric data points used here are summarized in Table~\ref{tab:mag}.

Following \citet{morishita19}, we simultaneously fit these photometric+spectroscopic data points by using synthetic spectral templates. We generate the fitting templates by using {\tt fsps} \citep{conroy09fsps}, with a setup of the MILES spectral library with the \citet{salpeter55} initial mass function, age pixels of [0.01,0.03,0.1,0.3,0.5,0.7,1.0,1.5,2.0,3.0]\,/\,Gyr, and metallicity in a range of $\log Z_{\odot} \in [-0.8,0.6]$ with a step size of 0.05, and the \citet{calzetti00} dust attenuation model. We calculate FIR dust emission templates based on the recipe provided by \citet{draine07}. Due to the limited number of FIR data points, we limit the calculation to their LMC model with $q_{\rm PAH}=2.37\%$.

In Figure~\ref{fig:sed}, we show the best-fit SED and star formation history of \id. The result indicates that \id\ had the last primary star formation activity $\sim0.5$\,Gyr ago, and then rapidly declined its star formation rate $\sim100$\,Myr before the observed redshift. Such a star formation history is in fact speculated from the absence of significant flux at rest-frame UV wavelength, as well as deep 4000\,\AA\ break and Balmer absorption lines seen in the deep grism spectra, characterizing \id\ as a post-starburst galaxy \citep{dressler83}. Such rapid decrease in star formation activity is also seen in other massive galaxies \citep{belli19,morishita19}, while their primary quenching process remains unclear. The low gas fraction and the inferred dust mass of the system found in this study potentially add further constraints on the primary quenching mechanism (Section~\ref{sec:dis}).

It is worth noting that the derived metallicity of \id, $\logZ \sim -0.2$\,dex, is $\sim0.4$\,dex below the mean relation of massive, non-compact quenched galaxies at $z\sim2$ \citep{morishita19}. While the statistical significance is still small due to the scatter around the mean relation ($\Delta\sim0.2$\,dex), this may further provide us a hint to its following evolution; if \id\ is representative of the compact galaxy population at this redshift, then a similarly low metallicity trend may be seen in local compact galaxies. Such a low metallicity implies that \id\ presumably experienced star formation of a short timescale in a relatively pristine gas environment \citep[e.g.,][see also below]{kriek16}.

The best-fit result returns dust mass $\log M_{\rm dust}/M_\odot=8.3\pm{0.6}$. {Despite the non-detection in ALMA Band 5, its strong upper limit constrains the FIR component sufficiently well within our providing templates, in combination with the MIPS $24\mu$m data point that constrains the polycyclic aromatic hydrocarbon (PAH) feature.} 
The best-fit dust mass derives dust-to-stellar mass fraction $M_{\rm dust}/M_* \sim 1.4\times10^{-3}$ {and $3\,\sigma$ upper limit on gas-to-dust ratio of $\simlt30$.} The inferred dust-to-stellar mass fraction is much higher than those of local early-type galaxies, ranging from $M_{\rm dust}/M_*\sim10^{-6}$ to $10^{-5}$ \citep{smith12,lianou16}, while a similar value was found from the stacking analysis of massive galaxies at $z\sim1.8$ in \citet{gobat18}. Such a significant amount of dust may play a critical role in suppressing star-formation activity, by preventing formation of hydrogen molecules \citep[e.g.,][]{kajisawa15,conroy15}. Our finding of the low molecular gas mass fraction derived above is consistent with this scenario.

Our estimate of dust mass is robust to different assumptions in the SED modeling. We also fit our photometric data points with another SED fitting code, {\tt piXedfit} \citep{abdurrouf21}\footnote{\url{https://github.com/aabdurrouf/piXedfit}}, which adopts broader parameter ranges for the dust component. The code returns $\log M_{\rm dust}/M_\odot = 8.2_{-0.6}^{+0.4}$, in good agreement with the dust mass estimated by {\tt gsf}.

No X-ray counterpart is found in the Chandra 7M\,sec catalog \citep{luo17} at the positions of \id\ or the tentative line in Sec~\ref{sssec:damato}. The non-detection is reasonable given its passively evolving nature inferred from the SED and undisturbed morphology.

%##########################################
\section{Discussion and Summary}\label{sec:dis}
In this study, we investigated the vicinity of a massive compact galaxy, \id, for molecular gas that can induce star formation and lead to strong size evolution to the local size-mass relation. Our two independent algorithms on the unique data cubes did not confidently detect any gas clumps or continuum emission, placing a conservative upper limit on molecular hydrogen mass.  
{We use this upper limit to further advance the argument presented in \citet{morishita16}, where they consider possible size evolution of \id\ through accretion of photometrically identified 34\,satellite galaxies. We here assume that each of the satellites 1.~has the smaller of $f_{\rm gas} \times M_*$ and $7\times10^9\,M_\odot$, 2.~converts gas to stars at $100\,\%$ efficiency, 3.~accretes to the central galaxy. The total mass contribution under this extreme assumption would be $\logm\sim9.5$, which still remains negligible compared to the observed mass of \id\ and the total stellar mass of the satellites ($\logm\sim10.8$). 
%\id\ is able to increase its size by a factor of $\sim5$ at most, which is still $\sim2\sigma$ below the local size-mass relation of early-type galaxies at $z\sim0$ \citep[e.g.,][]{shen03,taylor10,poggianti13}. 
The result secures the conclusion of \citet[][]{morishita16} that \id\ is unlikely to be on the size relation of the early-type galaxies, at least via minor mergers, ending up $\sim2\sigma$ below the local size-mass relation of early-type galaxies at $z\sim0$ \citep[e.g.,][]{shen03,taylor10,poggianti13} at most.
Since most of their satellite galaxies are not spectroscopically confirmed, this extrapolation is rather optimistic and the actual size evolution can be even less significant.}

In the local universe, on the other hand, there are a portion of compact, passively evolving galaxies in {\it high-density} regions i.e. in galaxy clusters \citep[e.g.,][]{valentinuzzi10}. These local compact galaxies, characterized by a similar light profile as for high-$z$ compact galaxies, dominate a significant fraction of cluster member galaxies at  $>3 \times 10^{10}M_\odot$, $\sim22\%$ \citep{valentinuzzi10}, whereas only $\sim4.4\%$ in general fields \citep{poggianti13}. These findings imply insignificant size evolution of (at least) some of compact galaxies at high redshift, as is found in simulations \citep[e.g.,][]{wellons16}. \citet{poggianti13} also found that compact galaxies in dense environments consist of older stellar populations than those in general fields. 

From these findings above, it is implied systematically earlier evolution of compact galaxies in dense fields, which is suggested by \citet{morishita17} and \citet{abramson18} from their structural analysis on cluster and field galaxies, as well as by direct comparison of stellar age of massive galaxies at $z\sim2$ \citep{wu18,carpenter20}. In fact, the stellar metallicity of \id\ implied from our SED fitting analysis is $\sim0.4$\,dex below the average trend of non-compact galaxies at the same redshift. Assuming that \id\ is representative of the progenitor population of local compact galaxies, then we may see systematical differences in their metallicity and chemical composition from those of other non-compact galaxies. While \citet{taylor10} found no significant metallicity offset of compact galaxies in general fields, currently such systematic comparison of local compact galaxies in dense environments has not been established.

Lastly, we revisit possible quenching mechanisms that could occur in \id. Our observations found an extremely low gas mass fraction in the system, $f_{\rm gas}\simlt\gf\%$, whereas the fraction ranges from $\sim50\%$ to $\sim100\%$ for the main sequence star-forming galaxies at similar redshifts \citep{daddi10,tacconi18,hayashi18b}. Such a low gas mass fraction of passive galaxies is not a surprise. For example, \citet{sargent15} reported a $3\,\sigma$ upper limit of $f_{\rm gas}\simlt5.8\%$ in a passively evolving galaxy at $z=1.43$. \citet{bezanson19} reported a $3\,\sigma$ upper limit $f_{\rm gas}\simlt7\%$ in a galaxy at $z=1.522$. Furthermore, the extremely low gas fraction of \id\ is worth comparing with those of compact galaxies {\it at an earlier phase}, or blue nuggets \citep{barro13,williams14,vandokkum15b}. \citet{barro17b} reported a short depletion time ($\sim27$\,Myr) of a star-forming compact galaxy at $z=2.3$ from their CO observations with ALMA. The galaxy is intensively forming stars at $\sim500\,M_\odot$\,/\,yr, which is somewhat comparable to the peak star formation rate of \id\ (Figure~\ref{fig:sed}).\footnote{Our SFR estimates are time-averaged at each pixel, and actual values could be higher if star formation activity is more instantaneous.} From this perspective, gas depletion by the past intense star formation activity can reasonable be considered as the primary cause of quenching for \id. 

We did not find gas clumps, disturbed gas structure, or dust emission in and around \id. Given that the last primary star forming activity occurred relatively recently ($\sim0.5$\,Gyr), this suggests that the past star formation activity was at least not caused by a galaxy-galaxy scale merger, and that gas was consumed inside the system rather than being ejected. This is consistent with a scenario derived from the local galaxies using stellar metallicity as an indicator \citep{peng15}. It is yet unclear what caused such intense star formation from our study. For example, \citet{talia18} observed an AGN-hosting compact star forming galaxy, and concluded that the star formation is likely caused by positive feedback from the AGN activity. This positive feedback scenario seems possible given a higher fraction of compact star forming galaxies host AGN than the overall star-forming population \citep[e.g.,][]{kocevski17,wisnioski18}. On the other hand, such high efficiency in star formation can also be induced without AGN activity \citep{dekel14,semenov18}, leaving the conclusion still pending. 

Given that the scenarios above are derived from one galaxy, we are still far from getting a general consensus on primary quenching mechanisms in compact massive galaxies. Nonetheless, in this study we showed that even non-detection of molecular gas and dust emission, in combination with a wide wavelength data coverage, sheds light on the nature of high-$z$ passive galaxies. Application of this new approach to archival data will immediately improve statistical arguments.

%=============================
\begin{deluxetable*}{ccccccccccc}%[!h]
\tabletypesize{\footnotesize}
\tabcolsep=3pt
\tablecolumns{11}
\tablewidth{0pt} 
\tablecaption{Photometric fluxes of \id, in units of $\mu$Jy.}
\tablehead{
\colhead{F125W} & \colhead{F140W} & \colhead{F160W} & \colhead{F225W} & \colhead{F275W}  & \colhead{F336W}  & \colhead{F435W}  & \colhead{F606W} & \colhead{F775W} & \colhead{F814W} & \colhead{F850LP}\\ 
\colhead{$K_S$} & \colhead{IRAC CH1} & \colhead{IRAC CH2}  & \colhead{IRAC CH3}  & \colhead{IRAC CH4}  & \colhead{MIPS $24\mu$m} & \colhead{MIPS $70\mu$m}  & \colhead{PACS $100\mu$m} & \colhead{PACS $160\mu$m} & \colhead{ALMA Band 5} & \colhead{}
}
\startdata
$3.26 \pm 0.02$ & $4.49 \pm 0.03$ & $5.30 \pm 0.03$ & $<0.14$ & $<0.05$ & $<0.03$ & $<0.01$ & $0.11 \pm 0.01$ & $0.26 \pm 0.01$ & $0.35 \pm 0.01$ & $0.62 \pm 0.02$\\
$7.99 \pm 0.09$ & $11.18 \pm 0.07$ & $11.98 \pm 0.06$ & $10.65 \pm 0.42$ & $6.09 \pm 0.44$ & $9.84 \pm 2.72$ & $<8538.11$ & $<2203.38$ & $<6610.15$ & $<21.40$ & 
\enddata
\tablecomments{
$1\sigma$ errors are quoted for those with $S/N>1$, and $1\sigma$ upper limits for the rest of the data points.
}
\label{tab:mag}
\end{deluxetable*}
%=============================

%##########################################
\section*{Acknowledgements}
{We thank the anonymous referee for providing constructive comments. This paper makes use of the following ALMA data: ADS/JAO.ALMA\#2019.1.01127.S. ALMA is a partnership of ESO (representing its member states), NSF (USA) and NINS (Japan), together with NRC (Canada), MOST and ASIAA (Taiwan), and KASI (Republic of Korea), in cooperation with the Republic of Chile. The Joint ALMA Observatory is operated by ESO, AUI/NRAO and NAOJ.}
T.M. is grateful to Kate Rowlands and Takuya Hashimoto for helpful discussion and advice on ALMA data reduction. Support for this work was provided by NASA through grant numbers HST-GO-15702.002, HST-GO-15702.002, and HST-AR-15804.002-A from the Space Telescope Science Institute, which is operated by AURA, Inc., under NASA contract NAS 5-26555.

{
{\it Software:} Astropy \citep{astropy:2013, astropy:2018}, numpy \citep{oliphant2006guide,van2011numpy}, python-fsps \citep{foreman14}, EMCEE \citep{foreman13}. }

%==================================================================
\bibliographystyle{apj}
\bibliography{/Users/tmorishita/adssample}
\end{document}